\documentstyle[aaspp4]{article}
\begin{document}

\title{Starburst-AGN Connection: A Lesson from High-$z$ Powerful
Radio Galaxies}

\author{Yoshiaki Taniguchi}

\affil{Astronomical Institute, Graduate School of Science, 
       Tohoku University, Aramaki, Aoba, Sendai 980-8578, Japan}

\begin{abstract}

Powerful radio galaxies at high redshift are highly useful in studies
of early evolution of AGN-hosting galaxies because their observed
optical and near infrared light are dominated by their stellar population
rather than the nonthermal continuum emitted by the central engine of AGNs.
In addition, the presence of AGN activity in them implies that a supermassive
black hole has been already made in their nuclei. These properties allow us 
to investigate a possible starburst-AGN connection in
early universe and then provide some crucial hints for the formation mechanism
of supermassive black holes. Taking observational properties of high-$z$
powerful radio galaxies into account, we discuss a possible formation
mechanism of supermassive black holes in their nuclei.

\end{abstract} 

\section{Introduction}

The formation and evolution of supermassive black holes were once thought
to be related only to some exotic phenomena such as nonthermal nuclear activity.
Active galactic nuclei (AGNs), in particular, quasars and powerful radio galaxies,
are basically minor populations among galactic nuclei. Therefore,
it has long been presumed that the history of supermassive black holes (SMBHs)
is independent from that of galaxies or that of stars. However, since the
discovery of a massive dark object in the heart of Andromeda galaxy (M31;
Dressler \& Richstone 1988; Kormendy 1988), it has been reported that 
a number of nearby normal galaxies appear to have such a massive dark object
in their nuclei (e.g., Kormendy \& Richstone 1995). 

Then, in the 90s, 
beautiful evidence for a SMBH was found in one of famous nearby AGNs,
NGC 4258, based on kinematic motion of H$_2$O masing clouds around the nucleus 
(Miyoshi et al. 1995).
Further challenge made in this decade has brought the discovery of a
tight correlation between the spheroidal mass and the SMBH mass 
(Magorrian et al. 1998; see also Silk \& Rees 1998).
 Surprisingly, normal galaxies and AGN-hosting galaxies
follow the same correlation  (Gebhardt et al. 2000; Ferrarese et al. 2001;
Wandel 2002). This suggests that most {\it nucleated} galaxies
have a massive dark object, a SMBH, despite of the nonthermal nuclear activity.
It is also suggested that AGN phenomena are not associated with
the formation process of a SMBH itself (e.g., Taniguchi 2002).
More importantly, the formation of SMBHs may be physically linked to
that of spheroidal component of galaxies.
In order to explore this issue, it seems important to investigate 
young galaxies in early universe. For this purpose, we use observational properties
of high-$z$ powerful radio galaxies\footnote{The most distant HzPRG known to date
is TN J0924$-$2201 (van Breugel et al. 1999). However, most HzPRGs are 
located at $z \sim 2$ - 4 (e.g., McCarthy 1993).}
(HzPRGs) because their host galaxies
have been intensively investigated for these two decades (e.g., 
McCarthy 1993) and consider a possible scenario for the 
the formation of SMBHs at high redshift (see for review, Rees 1984; 
Haiman \& Quataert 2004). 
 
\section{Nature of High-$z$ Powerful Radio Galaxies}

HzPRGs have been often used to
investigate stellar populations in young galaxies because their
optical and NIR continuum emission is dominated by host stars
rather than the nonthermal continuum emitted from the central engine
of AGNs (see, for a review, McCarthy 1993; see also Iwamuro et al. 2003
and references therein). Another reason why HzPRGs are useful in 
investigations of starburst-AGN connection in early universe is that
their host galaxies appear to experience of passive evolution.
Since their hosts are basically massive populations (i.e., their local
analogue are giant elliptical galaxies), they can be used to
investigate possible early connection between vigorous star formation
activity and early formation of a SMBH.

We summarize several lines of evidence for the passive evolution
of HzPRG host galaxies. First evidence comes from evolution of
optical/NIR spectral energy distributions (SEDs) of radio galaxies from
high redshift to the present day. In Figure 1, we show the SED evolution
of radio galaxies analyzed by Yoshii et al. (1994). It is shown that
host galaxies at high redshift tend to have blue SEDs (i.e., dominated by
massive young stars) while those in nearby universe tend to have
red ones (i.e., dominated by cool, old stars). This trend suggests 
strongly that hosts of radio galaxies have experienced the passive evolution.
The second evidence is that the submm (850$\mu$m) luminosity of radio 
galaxies shows a monotonic decrease with decreasing redshift from $z \sim 4$
to $z \sim 0$ (Archibald et al. 2001).
This property seems consistent with a scenario that
the first major epoch of star formation in massive hosts of radio galaxies
occurred at similar redshift and then evolved passively without
another episodic massive starburst during the course of their evolution.
The third evidence is that a number of HzPRGs are very luminous in FIR and/or submm
(e.g., $L_{\rm FIR} > 10^{12} L_\odot$, suggesting that very luminous
starbursts occurred in the gas-rich hosts (Archibald et al. 2001;
De Breuck et al. 2003 and references therein).
This also suggests that HzPRGs are basically similar to ultraluminous
infrared galaxies in the local universe that experience {\it
dissipative collapse} in their hosts (Kormendy \& Sanders 1992).
The fourth evidence comes from the probable superwind activity in
HzPRGs (e.g., Motohara et al. 2000; Taniguchi et al. 2001;
see also De Breuck et al. 2000)
because such large-scale superwinds could be attributed to a
very luminous starburst in their hosts (Larson 1974; Arimoto \& Yoshii 1987).
Since all these results are obtained by using fairly large samples of 
radio galaxies, one may conclude that hosts of {\it most} radio galaxies 
were very massive even at such high redshift and then
experienced the passive evolution after their initial starburst.

\begin{figure}
\epsfysize=10cm \epsfbox{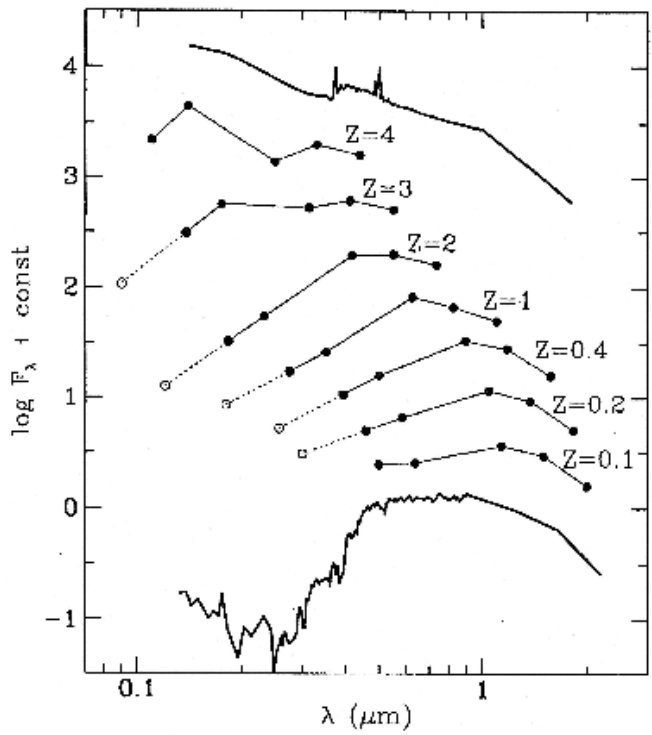}
\caption[]{
The evolution of optical/NIR spectral energy distributions of
HzPRGs (Yoshii et al. 1994). The ordinate is in arbitrary 
units. The upper SED shows that of nearby Irregular galaxies while
the lower SED shows that of nearby ellipticals.
\label{fig1}
}
\end{figure}

The above argument seems to be inconsistent with hierarchical 
clustering scenarios that have been accepted as a reliable mass
assembly history (e.g., Kauffmann et al. 1999). 
However, recent submm surveys carried out with SCUBA on JCMT 
have found a number of massive galaxies either with ultra (or hyper) 
luminous starbursts or AGN at high redshift. Any current hierarchical models 
underpredict the number of such massive galaxies in early universe
(e.g., Genzel et al. 2004). Therefore, it turns out that some massive 
galaxies were already assembled at high redshift and then evolved
passively to the present day.
If this is the case, it is expected that hosts of HzPRGs were born 
at very high redshift (e.g., $z \sim 10$) and most stars were formed
within a short duration in their early phase. After this initial starburst,
through the superwind phase, 
they passively evolved to the present-day massive giant elliptical galaxies
in which the Magorrian relation is well established.
If the formation of SMBHs in these galaxies wes tightly linked to
the initial starburst event, it seems worth studying which 
physical mechanism played an important role in their initial starburst phase.

\section{Possible Scenario for the Early Formation of SMBHs}

As discussed in the previous section, 
the observational properties of HzPRGs suggest that they could evolve
to giant elliptical galaxies in the local universe. 
Since giant ellipticals follow the Magorrian relation, the following questions arise as;
(i) {\it When was the Magorrian relation established ?}, and (ii) {\it
How was the Magorrian relation established ?}. 
In order to give some answers to the above questions, we consider probable
star formation history in the hosts of HzPRGs.

First, let us discuss the epoch of major star formation the HzPRG hosts.
An important observational result is that many HzPRGs show Nitrogen 
overabundance in their ionized gaseous nebulae (Vernet et al. 2001).
This property is also found in the broad line gas in high-$z$ quasars
(Hamann \& Ferland 1993). As for high-$z$ quasars, it is also known that
Fe is more abundant with respect to $\alpha$ elements like Mg
(Kawara et al. 1996; Dietrich et al. 2003).
The N overabundance could be attributed to
mass loss phenomena from intermediate mass stars.
The Fe overabundance could be attributed to much contribution
of Type Ia supernovae rather than Type II ones.
Therefore these observations suggest that the epoch of major
star formation could occur $\sim 1$ Gyr before the observed epoch.
Since typical redshifts of HzPRGs and high-$z$ quasars are $z \sim 2$ - 4,
the initial starburst could occur at $z \sim 5$ - 10
(Hamann \& Ferland 1993; Yoshii et al. 1998). 

Compact remnants, such as neutron stars and stellar-sized black holes,
could be left after the death of massive stars formed in the initial starbursts.
Mergers among such compact remnants could help in the mass growth of
black holes (e.g., Taniguchi et al. 1999 and references therein). However,  
prior going to details of such mergers, it seems better to consider 
a possibility that much earlier formation of seed black holes because pre-galactic,
Population III (Pop III) objects could form at very higher redshift, 
$z \sim 20$ - 30\footnote{Pop III objects could be born earlier than the cosmic
reionization epoch, $z_{\rm reion} \sim 17$ (Spergel et al. 2003)}.

Recent theoretical investigations of Pop III objects have suggested that 
very massive stars could be born at the Pop III era, $M \sim 10^3 M_\odot$
(e.g., Bromm, Coppi, \& Larson 2002; Abel, Bryan, \& Norman 2002;
Nakamura \& Umemura 2003). If the stellar mass
exceeds 260 $M_\odot$, such a very massive star could leave a compact 
remnant with mass of $\sim 100 M_\odot$ (Heger et al. 2003); i.e., 
the formation of an 
intermediate-mass black hole (IMBH; Taniguchi et al. 2000). 
If this is the case, such an IMBH could work as a first seed black hole,
leading to the formation of a SMBH (Madau et al. 2004). 
Such very massive stars could be born in a dark matter halo with mass
of $\sim 10^6 M_\odot$ at most (e., Fuller \& Couchman 2000; Yoshida et al. 2003).
Since the star formation efficiency could be low, it is unlikely that
the growth of an IMBH can be made through successive mergers
among many very massive stars and/or compact remnants. This suggests that
the growth could be made by gas accretion onto an IMBH although 
the angular momentum transfer problem may exist.

The gas accretion (Eddington) timescale is estimated as 
$T_{\rm Edd} \sim 4 \times 10^7 (\eta/0.1) (L_{\rm Edd}/L)$ yr,
where $\eta$ is the radiation efficiency in rest-mass units,
given that the gas accretion occurs at Eddington-luminosity limit
(e.g., Krolik 1999). Let us assume that an IMBH was formed at
$z_{\rm IMBH}  = 25$. As stated before, typical redshifts of HzPRGs and 
high-$z$ quasars are $z \sim 2$ - 4, and thus it is suggested that 
the initial starburst could occur around $z \sim 5$ - 10 in their hosts.
For simplicity, we adopt the epoch of initial starburst in a host of 
HzPRGs, $z_{\rm initial} = 8$ and that of AGN activity, $z_{\rm AGN} = 3$.
Given  a flat universe with $\Omega_{\rm matter} = 0.3$,
$\Omega_{\Lambda} = 0.7$, and $h_{70}=1$ where $h_{70} =
H_0/($70 km s$^{-1}$ Mpc$^{-1}$) based on the 
Wilkinson Microwave Anisotropy Probe (WMAP) Observations
(Spergel et al. 2003), the time interval between
$z_{\rm IMBH}$ and $z_{\rm initial}$, $\Delta T \simeq 6 \times 10^8$ yr.
This time interval corresponds to $\sim 15 T_{\rm Edd}$. Therefore, 
if the Eddington-limited gas accretion could occur, an IMBH could grow
up to the mass of $M_\bullet \sim 10^5 M_{\rm IMBH} \sim 10^7 M_\odot$
at $z = 8$. At this redshift, successive mergers among dark matter mini halos
could give rise to the formation of massive gaseous systems with
$M \sim 10^{11} M_\odot$. If a few or several such gaseous systems 
could merge into one, it is expected that a very luminous starburst
could occur near the merger center; note that ultraluminous infrared 
galaxies are local analogs of such merger-driven starbursts
(e.g., Kormendy \& Sanders 1992; see also Taniguchi, Ikeuchi, \& Shioya 1999). 
Such merger-driven luminous starbursts can be caused by strong dynamical
action of a binary or multiple SMBH system (Taniguchi \& Wada 1996;
Taniguchi \& Shioya 1998). A large number of super star clusters 
could be formed in the central region of the merging system and they 
are expected to sink into the merger center\footnote{Runaway mergers in some
super star clusters may lead to the formation of IMBHs (e.g., Mouri \&
Taniguchi 2002)}, leading to the formation of
a SMBH with $M_\bullet \sim 10^9 M_\odot$ within 1 Gyr (Ebisuzaki et al. 2001).
Given the same flat universe, the time interval between $z=8$ and $z=3$
corresponds to 1.4 Gyr. Therefore, SMBHs can be made in the host galaxies of
HzPRGs at $z = 3$. 

\section{Concluding Remarks}

We summarize a possible scenario for the formation of a SMBH
in nuclei of HzPRGs (as well as quasars) at $z \sim 3$
(see also Table 1). 

\begin{description}

\item{Step I.} A very massive star (i.e., a Pop III object) could born
in a dark matter mini halo at $z \sim 25$. Shortly after the formation, an IMBH 
with $M_\bullet \sim 100 M_\odot$ could be left as a compact remnant;
i.e., the first-stage seed black hole.

\item{Step II.} The gas accretion at the Eddington limit could grow up
an IMBH to $M_\bullet \sim 10^7 M_\odot$  
at $z \sim 8$ during the course of successive
mergers among dark matter halos; this can be regarded as
a seed SMBH. This growth takes 0.6 Gyr from $z \sim 25$.

\item{Step III.} A few or several massive gaseous system with $M \sim 10^{11}
M_\odot$ could merge into one. In their final phase, a large number of super star
clusters could be born near the merger center because of strong dynamical 
action driven by binary of multiple seed SMBHs. 

\item{Step IV.} Such super star clusters could sink into the merger center because
of dynamical friction, leading to the formation of a SMBH with $M_\bullet \sim 10^9
M_\odot $at $z \sim 3$. This process takes $\sim$ 1 Gyr, being shorter than the time interval
between $z=8$ and $z=3$ given the currently accepted WMAP cosmology.

\end{description}

The above scenario should be related to the formation of a massive giant
elliptical galaxy. The very luminous starburst (or the initial starburst)
could contribute to the formation of the spheroidal component. Therefore,
the Magorrian relation could be established in the Step IV in this scenario.
It is noted that such merger-driven ultraluminous starbursts could explain 
the Magorrian value of $M_\bullet/M_{\rm spheroid} \sim 0.001$
(Bekki \& Couch 2001). 

%-----------------------------------------------------
%    Table 1
%-----------------------------------------------------
{
\scriptsize
\tablenum{1}
\tablewidth{6.5in}
\begin{deluxetable}{ccclc}
\tablecaption{%
A Summary of the Proposed Scenario for the formation of SMBHs
}
\tablehead{
   \colhead{$z$} &
   \colhead{$T_{\rm Univ}$ (Gyr)} &
   \colhead{$\Delta T$ (Gyr)} &
   \colhead{Event} &
   \colhead{$M_\bullet$ ($M_\odot$)} 
}
\startdata
25 & 0.35 &       &  Pop. III $\Rightarrow$ IMBH  & $\sim 10^2$  \nl
   &      & 0.55  &  gas accretion                &              \nl 
 8 & 0.90 &       &  onset of mergers among massive halos & $\sim 10^7$  \nl
   &      & 1.40  &  starburst $\Rightarrow$ super star clusters  & \nl
 3 & 2.30 &       &  sinking into the merger center & $\sim 10^9$  \nl
\enddata
\end{deluxetable}
}

%\vspace {0.5cm}

\section*{Acknowledgments}
The author would like to thank Satoru Ikeuchi, Hideaki Mouri, Yasuhiro Shioya,
and Piero Madau for useful discussion
on possible formation mechanisms of supermassive black holes at high redshift.
He also would like to thank Kazuo Makishima and Shin Mineshige for their
kind invitation to this fruitful conference.

%----------------------------------------------------------------------------
%      References 
%----------------------------------------------------------------------------

\end{document}